\documentclass{article}
\usepackage{subfigure}
\usepackage{siunitx}
\usepackage{caption}
\usepackage{amsmath,graphicx}
\usepackage[preprint]{spconf}
\usepackage{setspace}
\usepackage{verbatim}
\usepackage{xcolor}
\usepackage{algorithm}
\usepackage{algorithmic}
\usepackage[numbers]{natbib}
\setcitestyle{square}

\title{Short video-based advertisements Evaluation System: \\Self-Organizing Learning Approach}
\makeatletter
\def\@name{{Yunjie Zhang\textsuperscript{1, 2}}\sthanks{This work was done while the author was an intern at Kwai Y-tech lab.}, {Fei Tao\textsuperscript{1}} , {Xudong Liu\textsuperscript{3},Runze Su\textsuperscript{1, 4}}, Xiaorong Mei\textsuperscript{3}, \\Weicong Ding\textsuperscript{1}, Zhichen Zhao\textsuperscript{5}, Lei Yuan\textsuperscript{1}, Ji Liu\textsuperscript{1,5}\\}
\makeatother

\address{\textsuperscript{1}Seattle AI Lab, Kuaishou Technology\\\textsuperscript{2}Department of ECE, University of Texas at Dallas\\ \textsuperscript{3}Ads Platform, Kuaishou Technology\\ 
\textsuperscript{4}Department of Statistics and Probability, Michigan State University\\ \textsuperscript{5}AI platform, Kuaishou Technology}

\begin{document}
\maketitle
\copyrightnotice{\copyright\ IEEE 2021}
\toappear{To appear in {\it Proc.\ ICASSP 2021, 6-11 June 2021, Toronto, Ontario, Canada}}
\begin{abstract}

With the rising of short video apps, such as TikTok, Snapchat and Kwai, advertisement in short-term user-generated videos (UGVs) has become a trending form of advertising. Prediction of user behavior without specific user profile is required by advertisers, as they expect to acquire advertisement performance in advance in the scenario of cold start. Current recommender system do not take raw videos as input; additionally, most previous work of Multi-Modal Machine Learning may not deal with unconstrained videos like UGVs. In this paper, we proposed a novel end-to-end self-organizing framework for user behavior prediction. Our model is able to learn the optimal topology of neural network architecture, as well as optimal weights, through training data. We evaluate our proposed method on our in-house dataset. The experimental results reveal that our model achieves the best performance in all our experiments.
\end{abstract}
%ads in UGVs
%
\vspace{-3 mm}
\section{Introduction}
\vspace{-3 mm}
\label{sec:intro}
%video sharing apps TikiTok 去掉 换成 Youtube
As \emph{daily active users} (DAU) of video sharing apps, e.g, Youtube, Snapchat and Kwai, have rocketed in recent years, advertisers take advantage of this trend to promote their products or services through \emph{user-generated videos} (UGV)-based advertisements. Generally, user behavior-related metrics, e.g., \emph{click-through rate} (CTR)\footnote{CTR describes how many users become interested in product based on video content.}, 3-second play rate\footnote{3-second play rate describes how much the users are attracted by the content of the beginning three seconds.}, are employed to assess advertisement quality and performance. These two metrics can be calculated as follows: \emph{CTR = number of clicks\footnote{The event that an user click the link that comes with advertisement.} $\div$ impressions\footnote{Impression here refers to the count of the event that a video is fetched from dataset and recommended to users.} }; and \emph{3-second play rate = number of plays (more than 3s) $\div$ impressions.} Recent researches \cite{wang2019exploring,wang2019mmctr} on recommender system require user profiles, which are extracted from user browsing history and user basic information, as essential model input to make precise prediction on video CTR within each user account. However, in some application scenarios, such as cold start and automatic generation of advertisements, where user-related information cannot be obtained, advertisement publishers have to rely on video content to estimate advertisement performance. Accordingly, methods for making precise prediction on UGV-based advertisement performance without user profile are of great value. \emph{Multi-Modal Machine Learning} (MMML), which exploits signals of different modalities jointly, is able to help in mentioned video-related tasks.

 Application of MMML tasks has been widely studied, including emotion recognition \cite{tao2018ensemble, liu2018multi}, object localization \cite{arandjelovic2018objects, zhao2018sound}, speech recognition \cite{Tao_2018_2, afouras2018deep, Tao_2018_4}, speech separation \cite{wu2019time}, voice activity detection \cite{tao2019end, tao2020end} and etc. We notice that multi-modal fusion strategy plays a decisive role in multi-modal tasks. Previous works have proposed sophisticated fusion methods and achieved remarkable success. %Different from targets listed above, advertising performance is related to user behavior and UGV-based ads contain more complex and unconstrained information. 
 However, in our case, previous solutions have two limitations: 1) they mainly focus on signal perception-related tasks, rather than user behaviors; 2) it is unclear how modalities interact with each other. For example, in ASR task \cite{Tao_2018_2}, visual content is taken as auxiliary information for audio content and therefore, audio modality is taken as query information in attention model. However, in our case, we have no prior knowledge about the relationship between input modalities and use behaviors. These shortages may lead to difficulties of applying existing methods in computational advertising.  

Our study focus on building an end-to-end system for predicting CTR and 3-second play rate of UGV-based advertisements. The contribution of our work can be summarized as follows: 1) To the best of our knowledge, our proposed system is the first work of predicting CTR and 3-second play rate directly from video content (combining audio and visual modalities), which is equivalent to predicting user behavior directly from raw signals. 2) We propose a self-organizing system that is able to learn the optimal topology of neural network architecture. More specifically, it is a data-driven framework which can adjust information flow by changing model architecture. %We propose a self-organizing system that is able to adjust information flow by changing model architecture through training data. More specifically, our framework learns the optimal topology of architecture, in addition to the optimal weights, to satisfy advertising-related demands. 

%Additionally, the proposed method introduces metadata information into model learning. Users' potential reactions towards ads, e.g., clicking on the link or making purchase, depend not only on videos' quality but also on advertised products themselves, as users may show less interest in unaffordable products.

We evaluated our proposed method on a video dataset which consists of 9841 advertisements videos collected from Kwai, a trending short video app worldwide. All videos are uploaded by advertisers and contain unconstrained information. The experimental results of CTR and 3-second play rate prediction reveal that the proposed method outperforms all models for comparison. %achieved $\num{1.77e-05}$ and $\num{1.43e-03}$ of mean absolute error (MAE) respectively and outperforms baseline model and all-connected model. 

%The rest of this paper is organized as follows. Related work is introduced in Section II. Section III provides details of our method and describes its implementation. Section IV introduces dataset and Section V presents experimental results on predicting different ad performance. Conclusions are drawn in Section VI.

\vspace{-3 mm}
\section{Related Work}
\vspace{-3 mm}
\label{sec:related}
%1. 推荐系统挺好的
%第一个从视频学user behavior 我主要从recommender system 和MMML 获取了一切启发。knowledge.
To build the first framework for predicting user behavior from videos, we borrowed ideas from recommender system and MMML studies. Recommender system relied on designated data pre-processing to collect descriptive features. Google's Wide-\&-Deep model \cite{cheng2016wide}, which has been widely deployed in industry, combined these features at different levels within one neural network. Recently proposed AutoCTR framework \cite{song2020towards} explored the optimal model structure in a data-driven way. These ideas inspired us to design a model that is able to process features collected from different levels. However, AutoCTR and Wide-\&-Deep model may not be applicable in our task, as they could not handle raw signal inputs. \emph{Densenet} \cite{huang2017densely} essentially had similar strategy to Wide-\&-Deep model that it used cross-layer connection in image classification task. It merged information from different levels in \emph{dense block}, where one layer was directly connected to all its subsequent layers. 
%We saw Densenet as an advanced version of Wide-\&-Deep model.

%删掉第一句话 我们了解到 如何融合是个重要可以。better fusion strategy can help us in video-content modeling. 
%For MMML research fusion strategy is the key 为了更好就融合audio video stream
%We have mentioned that multi-modal fusion method plays a decisive role in video-related tasks. Generally, previous multi-modal fusion methods employ models with fixed architectures. 
In MMML research, one straight-forward fusion method was to combine weighted prediction results across all modalities, where the weight of each modality was determined by its own performance on validation set. This method may fail in handling UGVs, whose modalities had different significance across UGV topics. Another widely applied fusion method was to concatenate the features extracted from different modalities into a joint representation \cite{noroozi2017audio, afouras2018deep, wu2019time} and then the concatenated feature vectors could be processed by a classification/regression model. Such simple method has shown its effectiveness in many video-related tasks introduced in previous section. However, merging features into one vector did not provide enough flexibility in dealing with unconstrained videos. In many other works, attention model \cite{vaswani2017attention} has been considered  \cite{hu2019deep} to assign weights dynamically, where the strategy could be learned through end-to-end training. Also signal from one modality could be utilized as auxiliary information on other modalities \cite{Tao_2018_4, yu2020audio}. These methods made prior assumptions on relationship among all available modalities and set constrains on the architectures of fusion models, which was not the case in our study. To tackle the shortages mentioned above, we develop a self-organizing framework, which is able to explore the optimal topology of neural network architecture through training data. Our proposed method bridges these two research domains so that it is able to make prediction on user behavior with original video input.

\vspace{-3 mm}
\section{Dataset}
\vspace{-3 mm}
%做成5.1
In our study, we use our in-house dataset, containing advertisement play history within one week. %, which is collected from Kwai, a popular short video app from China.
The advertisements have been grouped into 19 pre-defined categories by advertisers. This video setting is same as KWAI-AD \cite{chen2020imram} dataset. All audio tracks are sampled 44.1kHz sampling rate and each audio track has two channels. We mix each track into mono-channel in this study. The visual tracks have the resolution of $720 \times 1280$, a typical vertical setup for mobile devices. All advertisements have the same \emph{frame per second} (FPS) of 25. Also, we summarize one-week performance for each advertisement, including impression, CTR and 3-second play rate. 

%我们只选择比较popular的广告，view 次数很多一天一万 广告更加有统计意义 These videos are good
%all videos are unique (but they may sama category)
However, CTR and 3-second play rate are lack of statistical significance without enough impressions. Therefor, based on our experience, we set 70,000 as impression threshold, under which advertisement samples have been discarded. After this step of filtering, a total of 9841 advertisements are collected. The total length of these advertisements is about 82 hours. %总时长 
%Additionally, we set an impression threshold, under which ads are discarded. As CTR is an extremely small value (usually less than 1\%), it is unreliable and meaningless without sufficient \emph{impressions}. Also, it has been noticed that some advertisers tend to re-used previous ad materials in new ads, indicating that duplicated videos exist in our dataset. We merge data according to the material similarity, especially image and text, and then, re-calculated performance based the merged statistics. After these two steps of data cleaning, a total of 9841 ads are collected. 

\vspace{-3 mm}
\section{Proposed Approaches}
\begin{figure}
    \begin{center}
    \includegraphics[width=0.55\textwidth,trim={3cm 2cm 0 1cm}, clip=true]{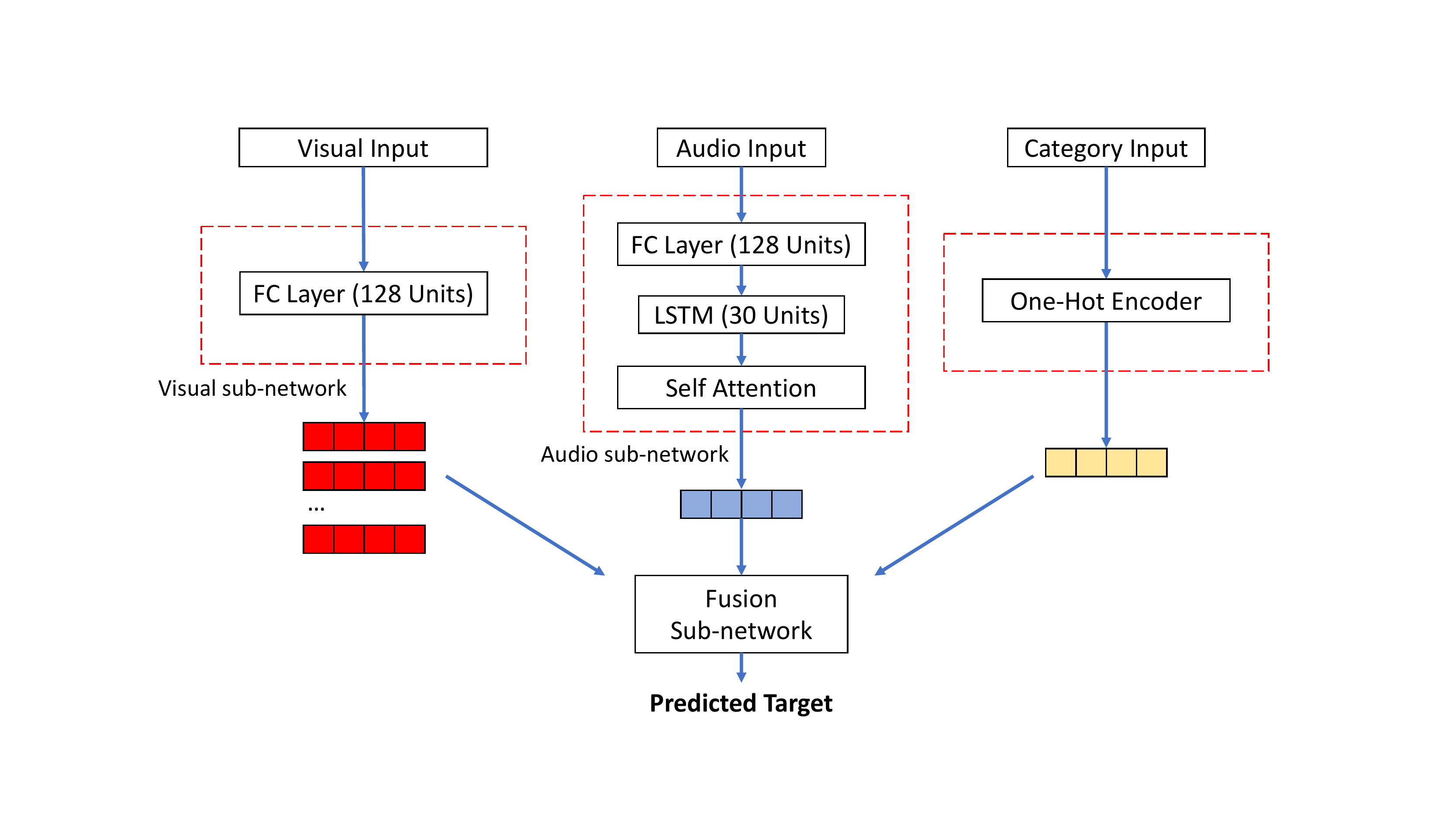}
    \vspace{-0.5cm}
    \caption{Overview of the proposed framework.}
    \label{fig:overview}
    \end{center}
    \vspace{-0.5cm}
\end{figure}

\vspace{-3 mm}
\label{sec:approaches}
The overview architecture of our system is shown in Figure \ref{fig:overview}. It consists of two parts: single modality sub-networks (visual and audio) and fusion sub-network. In this study, we train all these sub-networks jointly. 
%改一下
%强调一下end-to-end
%As mentioned before, instead of using a fixed architecture for fusion model, we deploy a self-organizing training procedure to explore the optimal topology.

\vspace{-3 mm}
\subsection{Feature extraction}
\vspace{-3 mm}
In our study, CTR is related to the entire video, while 3-second play rate only corresponds to the content in the beginning three seconds (based on the definition introduced in Section \ref{sec:intro}). Therefore, we prepare the feature extraction separately for these two tasks. For CTR prediction, we utilize our in-house key-frame extractor to extract 8 visual frames from each video and keep entire audio track. For 3-second play rate prediction, we extract 3 visual frames only from the first three seconds, one frame per second and only keep audio track of the first three seconds. 
%audio visual track
Extracted visual frames and audio tracks are then processed through Mobilenetv2 \cite{sandler2018mobilenetv2} and VGGish \cite{hershey2017cnn} respectively to generate visual and audio inputs. Our primary research showed that advertisement category also had impact on CTR prediction result and thus, we introduce category into our case as an extra modality. %our primary experiments show that ad category also has impact on prediction and thus, we introduce category into our case as an extra modality. 
Category information (we have 19 advertisement categories) are processed by an one-hot encoder to generate category embedding. %The embeddings are then repeated to match the frame count in each ad. 

\vspace{-3 mm}
\subsection{Single Modality Sub-networks}
\vspace{-3 mm}
%每一个layer 多少neuron 要写清楚 图上标的话 你要在图上说 neuron number for each layer are shown in XXX
%parameters are shared across all frames 不用单独说first dimension 拆开来讲 category input 这里也不用了
Inputs of visual and audio modalities are processed by two sub-networks respectively. The visual sub-network has a fully-connected (FC) layer, whose parameters are shared across all frames. The output of visual sub-network is a sequence of 128-D embeddings. The audio sub-network consists of a FC layer, an uni-directional LSTM layer and a self-attention layer. The output of audio sub-network is a 128-D embeddings. Numbers of neurons in each layer are shown in Figure \ref{fig:overview}. For each video, we have several frames for visual input, while we have only one embedding for audio input and one embedding for category. Therefore, audio embedding and category embedding are repeated to match match the frame count in each video. These collected embeddings are then sent to our proposed fusion model. 
%Since for each video we have multiple frame input, audio and category embeddings are repeated to match frame count so that they can be concatenated into one embedding matrix. 
%
\vspace{-3 mm}
\subsection{Fusion Sub-network}
\begin{figure}[tb]
        \centering
	\subfigure[Baseline model.]
	{
		\includegraphics[width=.45\columnwidth,clip=true, trim={10cm 3cm 11cm 2cm}]{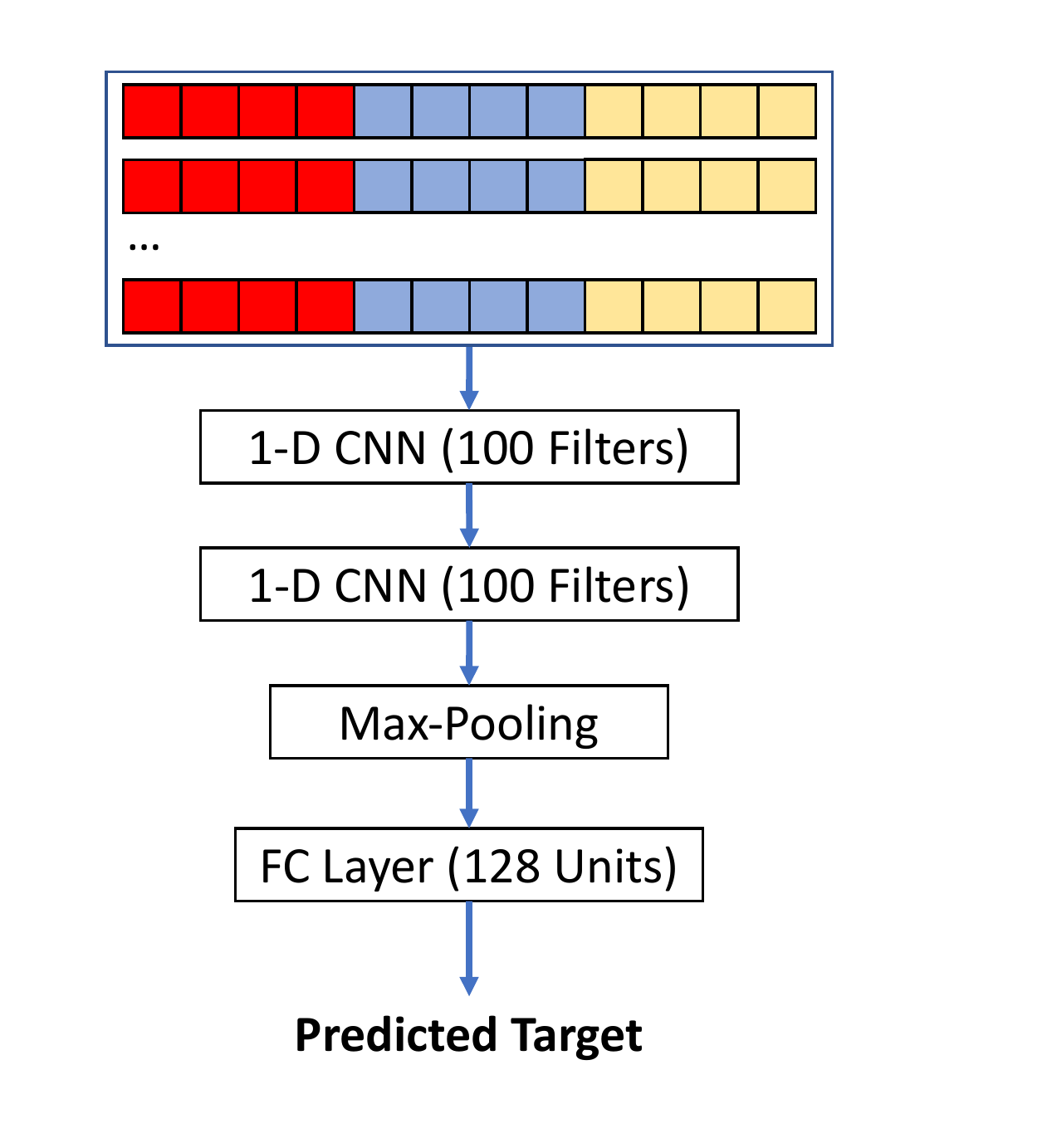}
        \label{fig:baseline-model}
	}
	\subfigure[All-connected model.]
	{
		\includegraphics[width=.45\columnwidth, clip=true, trim={10cm 3cm 11cm 2cm}]{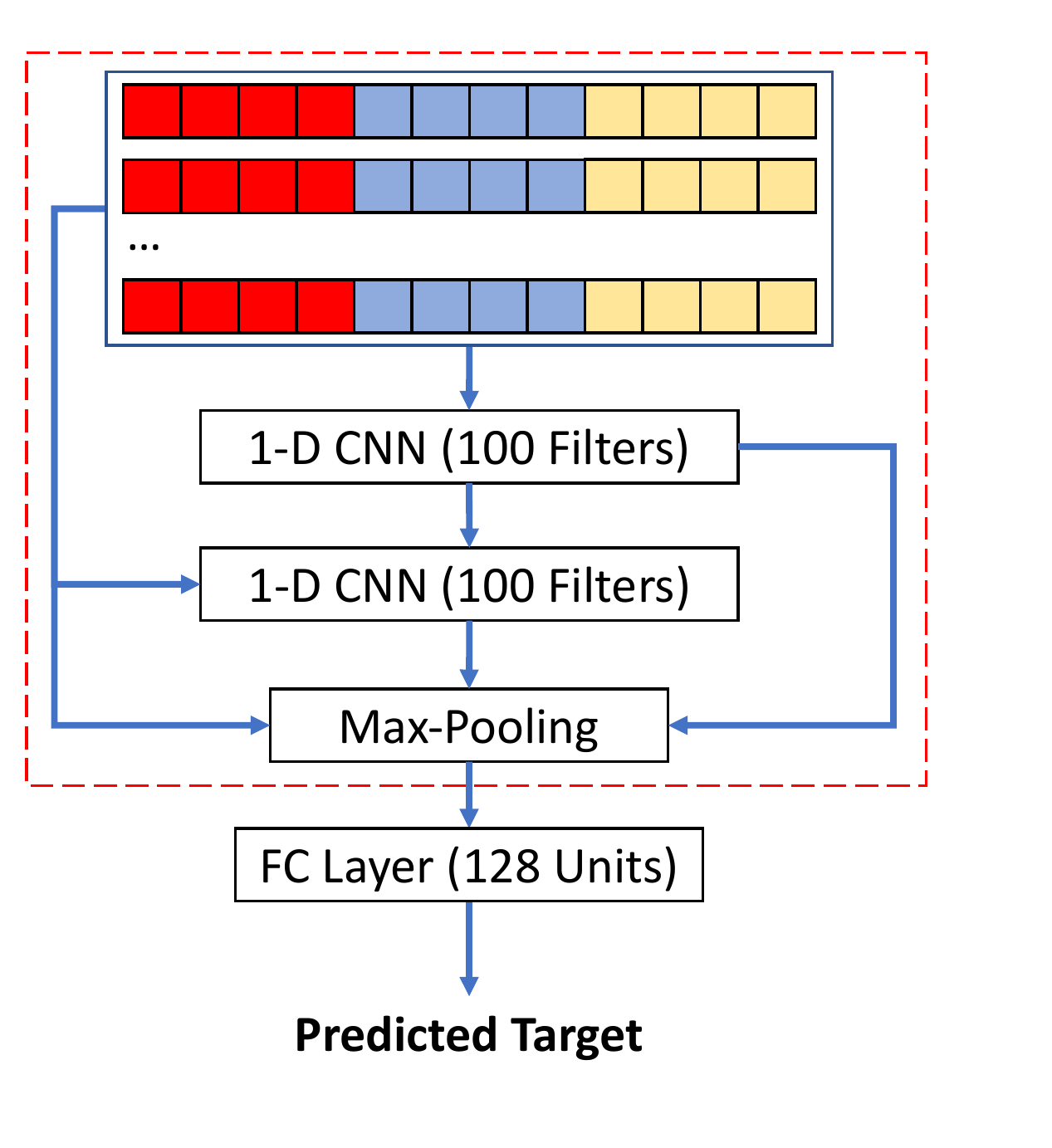}
        \label{fig:all-connected-model}
	}
	\vspace{-3mm}
	\caption{All-connected model is a modified version of baseline model. In the dashed border, each layer is connected to its subsequent layers.}
	\label{fig:models}
\end{figure}

\begin{figure}[tb]
    \centering
	\subfigure[Step (2)]
	{
		\includegraphics[width=.3\columnwidth,clip=true, trim={9.5cm 4.5cm 11.5cm 4.5cm}]{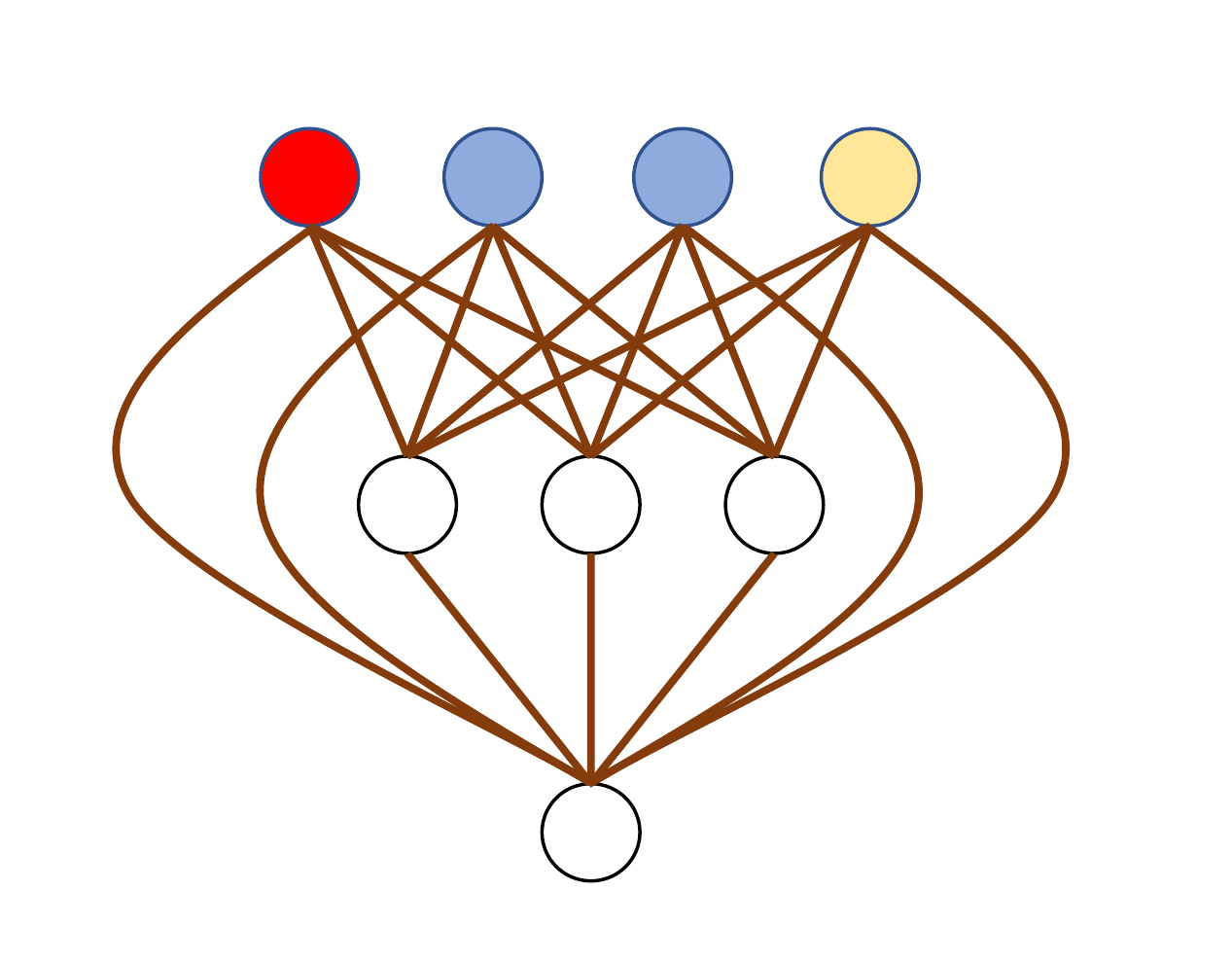}
        \label{fig:step-a}
	}
	\subfigure[Step (3)]
	{
		\includegraphics[width=.3\columnwidth, clip=true, trim={9.5cm 4.5cm 11.5cm 4.5cm}]{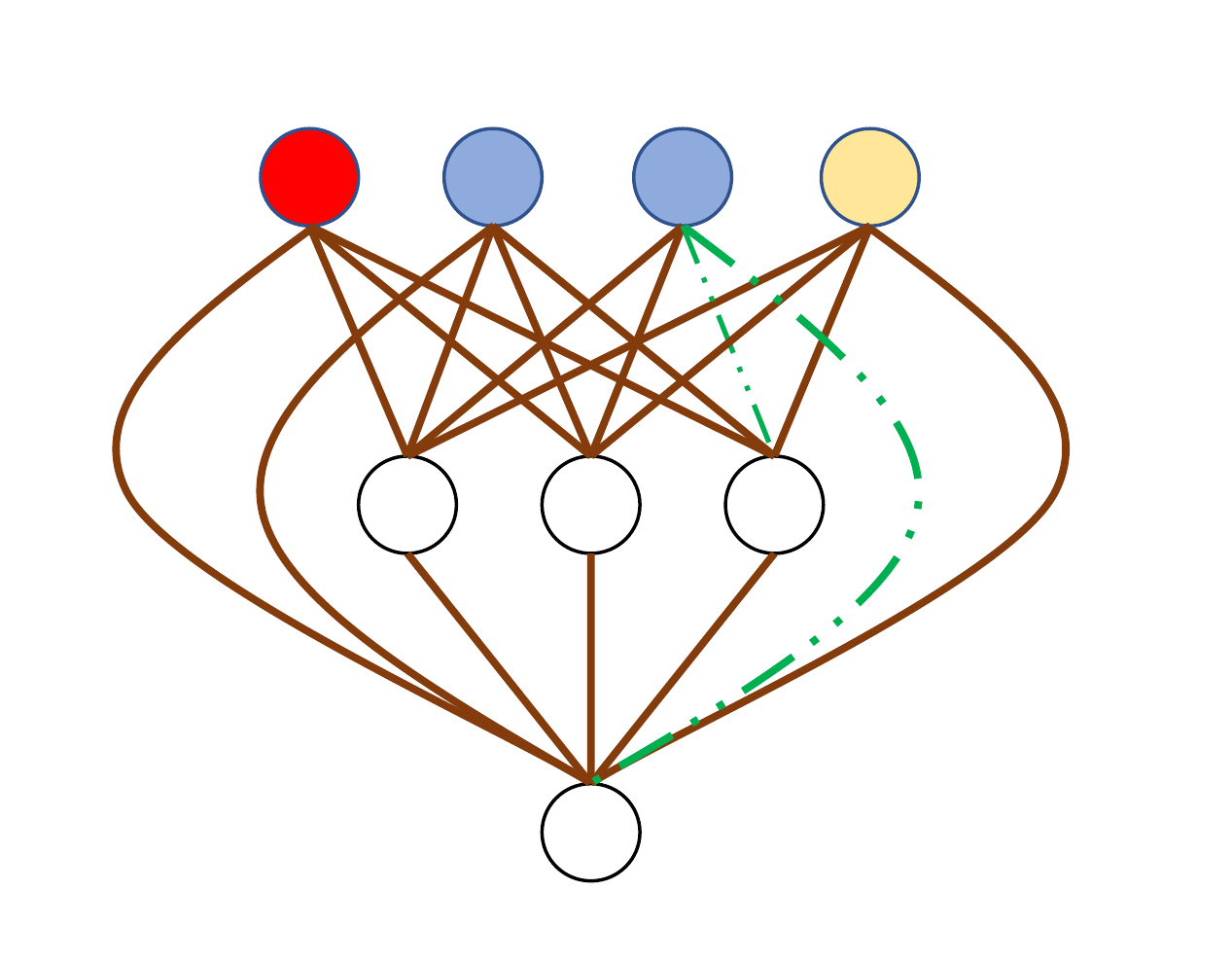}
        \label{fig:step-b}
	}
	\subfigure[Step (4)]
	{
		\includegraphics[width=.3\columnwidth, clip=true, trim={9.5cm 4.5cm 11.5cm 4.5cm}]{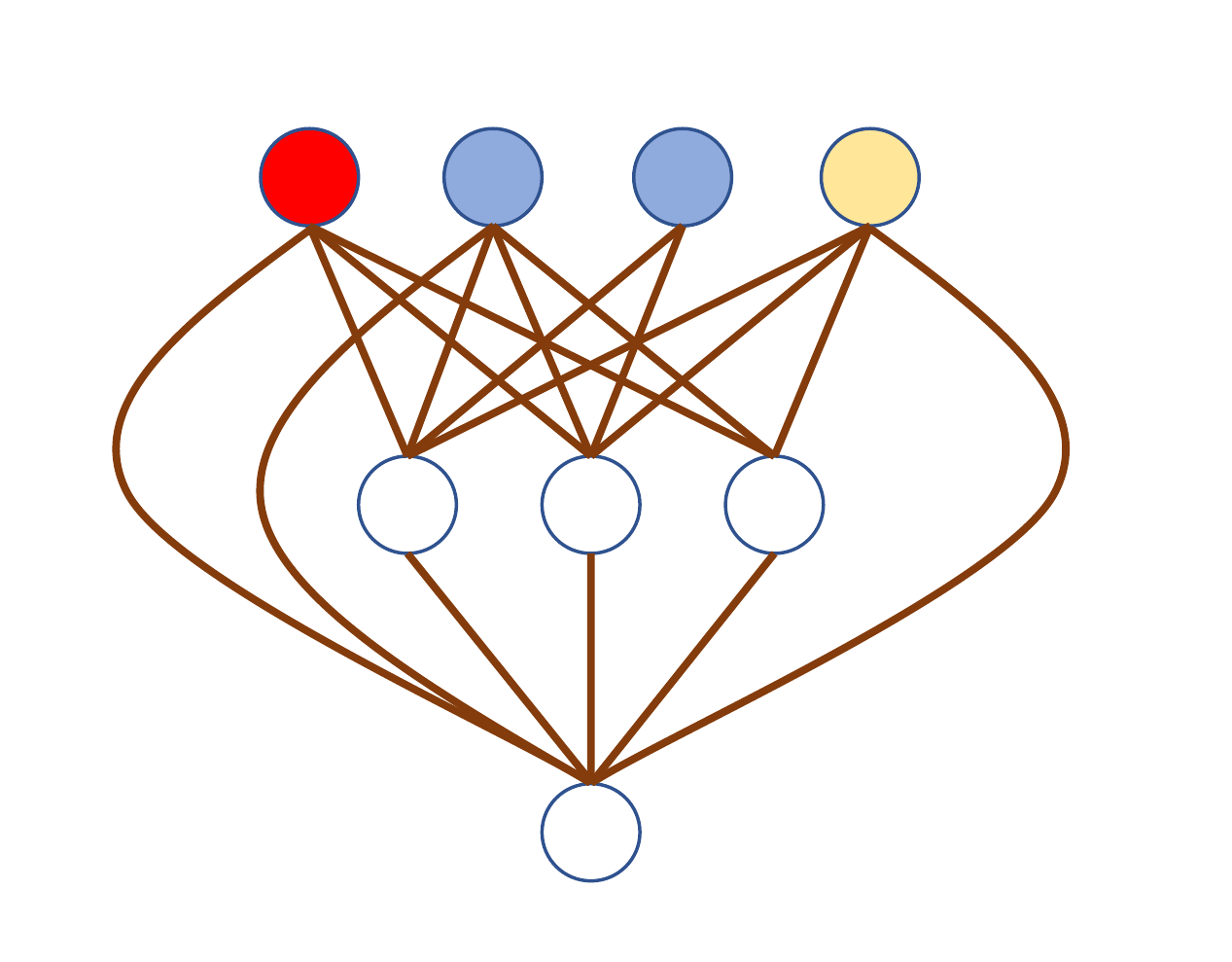}
        \label{fig:step-c}
	}
	\vspace{-3mm}
	\caption{Step (2), (3) and (4) of our self-learning approach. We simplify the diagrams for illustration.}
	\label{fig:steps}

\end{figure}

\vspace{-3 mm}
%数字统一一下
In our study, we have two fusion approaches: baseline and self-organizing. %the optimal topology of architecture is learned through self-organizing. name as "" 要么分号 要么句号 
For the baseline approach, we adopt fusion model sub-network proposed in Ti-AVC network \cite{su2020themes} (Figure \ref{fig:baseline-model}). It has two 1-D convolutional neural networks (CNN), one max-pooling layer and one FC layer to predict targets. For our proposed approach, which we name as ``self-organizing" approach, we follow the following steps to learn fusion strategy from data: (1) Modify the fusion sub-network in the baseline approach. We connect input embeddings to the second CNN layer and the max-pooling layer, in addition to the first CNN layer. The output of the first CNN layer is also connected to the max-pooling layer, as shown in dashed border of Figure \ref{fig:all-connected-model}. It is equivalent to connecting the output of each layer to \emph{all} of its following layers in the dashed boundary. Therefore, we name it as ``all-connected" fusion sub-network. (2) Optimize all-connected fusion sub-network until there is no more improvement in performance on validation set (shown in Figure \ref{fig:step-a}). (3) Select the 5\% of connections with the lowest absolute values (shown in Figure \ref{fig:step-b}). (4) Remove connections selected in step (3) and fine-tune the sub-network (shown in Figure \ref{fig:step-c}). (5) Repeat step (3) and step (4) until 
parameters number reaches a pre-defined threshold. We employ the parameter number of fusion model in baseline approach as our threshold in the experiments.

The logic behind removing connections is that connections with low absolute values indicate that they play less important role in forward propagation than others. With these connection removed, our model re-organizes information flow and learns the optimal topology. Therefore, we name our fusion model as self-organizing model. The entire procedure is data-driven and does not require manually defined rules. Our self-organizing framework is flexible and sophisticated. 

\vspace{-3 mm}
\section{Experiment and Analysis}
\vspace{-3 mm}
\label{sec:experiment}

%Ad owners create short videos to publicize their products and with the help of professional teams, these ads have considerably higher quality, compared with amateur-level UGVs. In real practice, it has been observed that data quality could be negatively affected by two factors. 1) As CTR is an extremely small value (usually less than 1\%), is unreliable and meaningless without sufficient \emph{impressions}. We note that impression here refers to the count of event that a video is fetched from dataset and recommended to users. Although Kwai's advertising system ensures that all ads can be exposed to users, the chance of exposure is not guaranteed to be consistent. (2) In order to boost procedure of ad producing, some advertisers tend to re-used previous ad materials in new ads. This could lead to ambiguity that two observably identical ads have significantly different performance, given inconsistent exposure strategy.

%In order to address mentioned issues, we adopt a two-step procedure to clean the dataset. First, we merge data according to the material similarity, especially image and text, and then, re-calculated performance based the merged statistics. After the step of deduplication, data items with low impressions are discarded so that all data has statistical significance. (I dont understand) A total of XXXX videos and their performance have been collected 14 days from they were created.  

\subsection{Experiment Setup}
\vspace{-3 mm}
We evaluated our proposed method on CTR prediction and 3-second play rate tasks. For each task, two types of experiments were conducted: regression and classification. For the classification experiment, we uniformly binned the data into five groups (every bin had same count of sample in training set), based on the distribution of data on training set (shown in Figure \ref{fig:ctr_distribution} and \ref{fig:3s_distribution}). Here, the 1-D output layer in regression model was replaced by a 5-D softmax output layer.

Two models, which have been introduced in Section \ref{sec:approaches},  were built for comparison with our proposed work. The first one was the fusion model in the baseline approach (named as ``Baseline"). The other model for comparison was the fusion model in all-connected approach (named as ``All-Connected"). It should be noted that ``All-Connected" model has more neuron connections than ``Baseline" model with the same neuron number. To make fair comparison, we trimmed its neuron number (51\% of kernels in convolutional layers and neurons in dense layer) to ensure it has same parameters number as ``Baseline" model. We also made the final parameters number of our proposed approach (named as ``Self-Organizing") same as the ``Baseline" model. In other words, all three models had the same number of parameters.

80\% and 10\% of the samples were randomly selected as our training set and validation set, while the rest were used as our testing set. We adopted \emph{Adam} as our optimizer. 0.0001 and 8 were chosen as learning rate and batch size respectively in all experiments. In all regression experiments, \emph{mean squared error} (MSE) and \emph{mean absolute error} (MAE) were employed as our evaluation metrics. To fairly compare all models' performance on different tasks, in addition to the two metrics above, we utilized ratio of MAE to Average ground truth (named as ``MAR" in our study) as our main evaluation metric. \emph{Mean absolute percentage error} (MAPE) was not used in our study, as it was more likely to be effected by outlier samples, which had low MAE but introduced extremely high MAPE. In classification experiments, the accuracy of all models were summarized. %Also, an early stopper is deployed in Baseline and All-Connected experiments to constantly monitor the validation loss. On-going training stage will be terminated, if validation loss keeps still for five epochs. %self-organizing 之前已经介绍过了

%To compare our proposed method with previous works, we build two models that have been introduced in section 3 for comparison. The first one is the baseline model, which employs cascade structure without any cross-layer connection. The fusion mechanism in the baseline model is just shallow concatenation of all three modalities in the first layer. The second model is all-connected model, in which one layer is connected to all its subsequent layers. However, to make the comparison fair, we trim 54\% of kernels in convolutional layers and neurons in dense layer to make total number of weights same as that of baseline model. Also, we setup a weight count threshold for our proposed model so that it will stop self-organizing once the weight count is lower than the threshold. As a result, all three models have similar number of weights. 
%it should be noticed that in the experiments, 
%all-connected(without any removing)
%我们的模型叫做 self-organizing
\subsection{Experiment Results and Analysis}
\begin{figure}[!b]
        \centering
    
	\subfigure[CTR distribution.]
	{
		\includegraphics[width=1.0\columnwidth,clip=true, trim={1cm 3cm 0 4cm}]{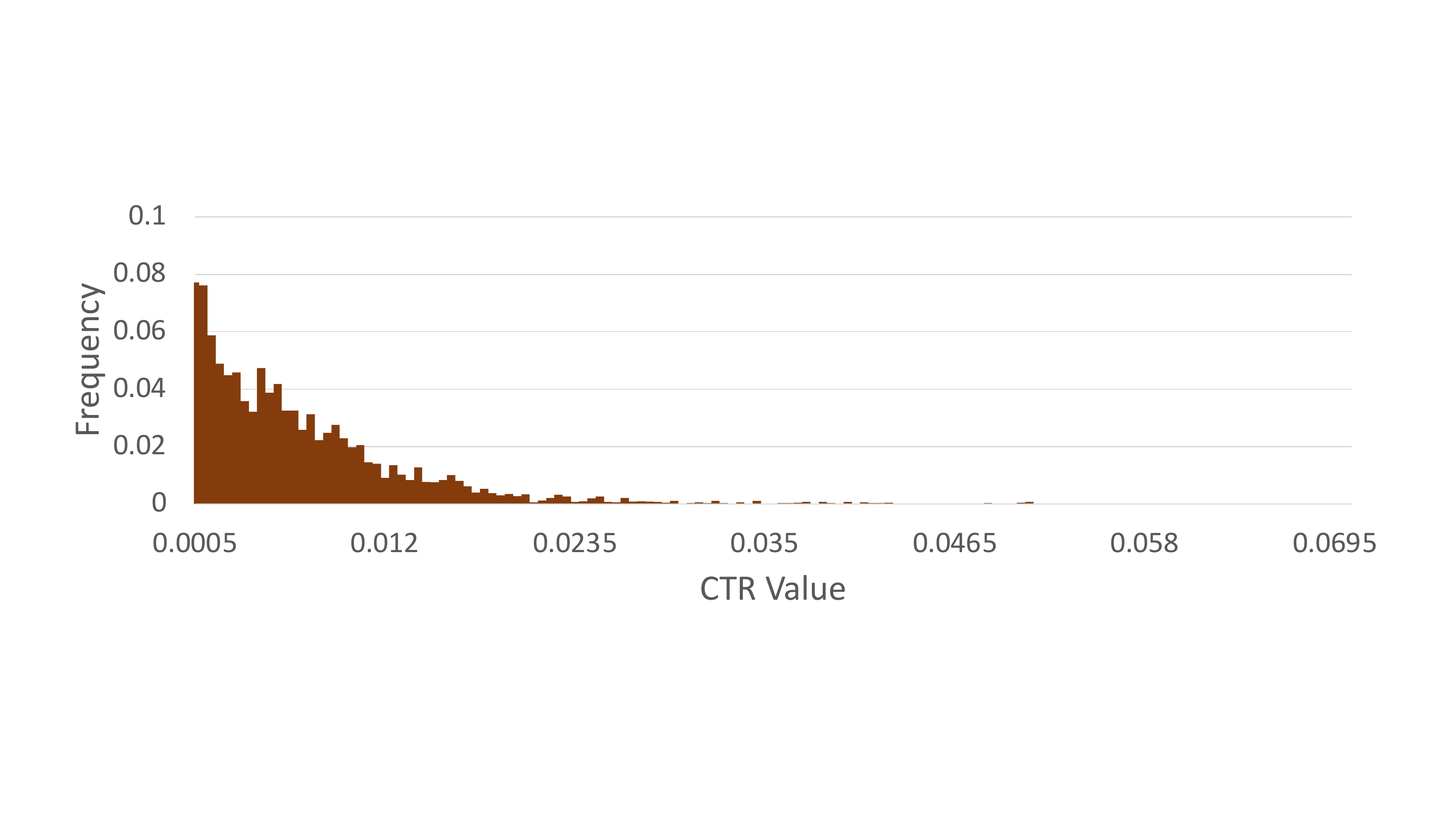}
	
        \label{fig:ctr_distribution}
	}
	\subfigure[3-second play rate distribution.]
	{
		\includegraphics[width=1.0\columnwidth, clip=true, trim={1.2cm 3cm 0 3.5cm}]{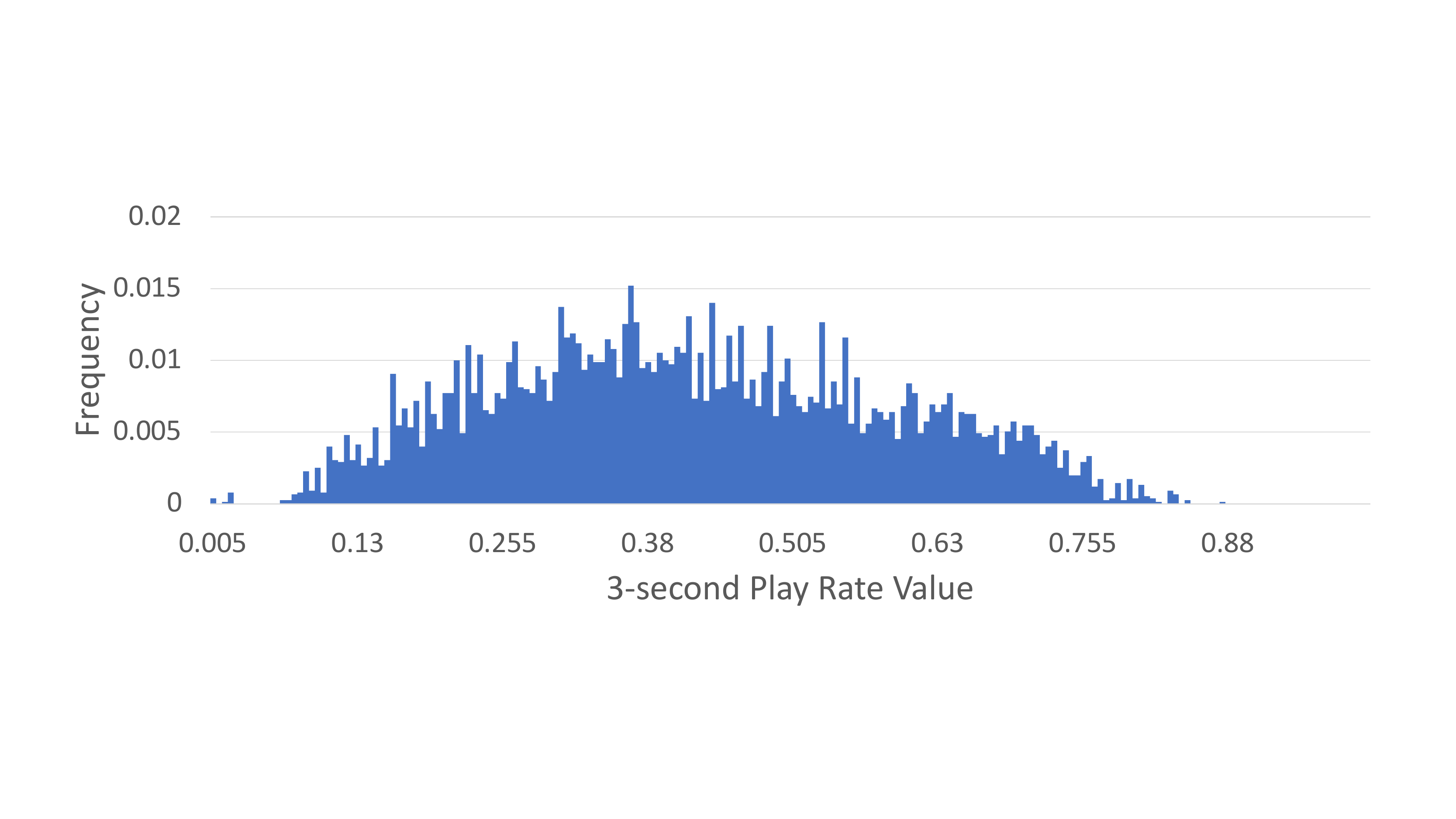}
        \label{fig:3s_distribution}
	}
	\vspace{-0.5cm}
	\caption{Training data distribution in regression tasks.}
	\label{fig:distribution}
\end{figure}

\vspace{-3 mm}
%这一段放上面
%For CTR ,our proposed method can achieve best performance We mainly focuso on MAR，因为别的太小了 没意义 outperform by (absolute difference). Our proposed method can achieve the best performance in both tasks. For 5-class classification (同样逻辑).
The results of all regression experiments are listed in Table \ref{tab:ctr_results} and \ref{tab:3s_results}. In CTR regression experiment, our proposed Self-Organizing model beats Baseline model and All-Connected model by 0.9\% and 6.0\% respectively (absolute difference). In 3-second play rate regression experiment, our proposed Self-Organizing model outperforms Baseline model and All-Connected model by 0.5\% and 0.3\% respectively (absolute difference). We notice that MAE of Self-Organizing model is 2.5\% lower than Baseline model and 1.8\% lower than All-Connected model. As shown in Figure \ref{fig:ctr_distribution} and \ref{fig:3s_distribution}, the CTR distribution follows an heavy-tail distribution, while 3-second play rate follows a normal distribution. In both types of distribution, our proposed model achieves the best performance, indicating that it has strong flexibility and generalization ability.

Table \ref{tab:results} summarizes classification experiment results. In CTR classification experiment, our proposed model outperforms Baseline model and All-Connected model by 5.0\% and 2.8\% respectively (absolute difference). In 3-second play rate classification experiment, our proposed model outperforms Baseline model and All-Connected model by 1.4\% and 1.3\% respectively (absolute difference). We note that classification is a task with courser granularity compared with regression. It shows that our proposed Self-Organizing model outperforms other baseline models in all granularities.%不知道这里有other baseline model好不好

\begin{table}[tb]
\centering
\begin{tabular}{l|p{1.6cm}|p{1.6cm}|p{1.3cm}}
\hline
Model          &  MSE & MAE & MAR \\
\hline
Baseline       &  $\num{1.87e-05}$   &$\num{2.26e-03}$ &   $35.5\%$   \\
All-Connected  &   $\num{2.00e-05}$ & $\num{2.58e-03}$ &   $40.6\%$   \\
\hline
\textbf{Self-Organizing} & $\mathbf{1.77\!\times\!10^{-5}}$&$\mathbf{2.20\!\times\!10^{-3}}$ &   $\mathbf{34.6}$\!\textbf{\%}  \\
%\multicolumn{1}{l|}{\textbf{1.77} $\times$ $\textbf{10}^{\textbf{-5}}$}
%$\mathbf{1.77\times10^{-5}}$
\hline
\end{tabular}
    \caption{Summary of experimental results of CTR prediction.}
    \label{tab:ctr_results}
\end{table}

\begin{table}[tb]
\centering
\begin{tabular}{l|p{1.6cm}|p{1.6cm}|p{1.3cm}}
\hline
Model          & MSE & MAE & MAR \\
\hline
Baseline       &  $0.0152$   &  $0.0744$   &   $17.6\%$   \\
All-Connected  &  $0.0148$   & $0.0738$ &   $17.4\%$   \\
\hline
\textbf{Self-Organizing} &  $\mathbf{0.0143}$   & $\mathbf{0.0725}$ &  $\mathbf{17.1}$\!\textbf{\%}   \\
\hline
\end{tabular}
    \caption{Summary of experimental results of 3-second play rate prediction.}
    \label{tab:3s_results}
\end{table}

\begin{table}[t!]
\centering
\begin{tabular}{l|c|l|c}
\hline
\multicolumn{2}{l|}{CTR}      & \multicolumn{2}{l}{3-second Play Rate} \\ \hline
Model          & Acc & Model             & Acc\\ \hline
Baseline       &      $61.3\%$         &  Baseline          &   $61.4\%$              \\ 
All-Connected  &     $63.5\%$ & All-Connected     &        $61.5\%$         \\ \hline
\textbf{Self-Organizing} &     $\mathbf{66.3}$\!\textbf{\%}          &  \textbf{Self-Organizing}    &     $\mathbf{62.8}$\!\textbf{\%}             \\ \hline
\end{tabular}
    \caption{Summary of experimental results of multi-class classification. ``Acc" here refers to accuracy.}
    \label{tab:results}
\end{table}

\section{Conclusion}
\vspace{-3 mm}
\label{sec:conclusion}
%in data driven fashion. 
%we proposed a self-learning approach which can predict use behavior directly 
%1.contribution两个点
%我们实验结果特别好
In this study, we propose a self-organizing approach, which can learn the optimal topology of neural network architecture in a data-driven way. Unlike previous approaches, our proposed method does not require prior knowledge or assumption about relationship among modalities. It provides more flexibility in handling tasks related to UGVs, which contain complex and complicated information. Also, our proposed method is able to predict CTR and 3-second play rate directly from video inputs. Our experimental results reveal that our proposed method successfully predict user behaviors and outperforms all other models for comparison.
%In this work, we explored the feasibility of self-organizing mechanism for multimodal fusion models. Unlike previous approaches, our proposed method does not require prior knowledge or assumption about relationship among modalities. It provides more flexibility in handling tasks related to short videos, which contains complex and complicated information. Also, our proposed method is able to predict CTRs and 3-second play rate, important statistics of user behavior in recommender system. We demonstrated our method on KWAI ads dateset, which consists of 9841 short video-based ads collected from advertisers. The experimental results reveals that our proposed method outperforms cascade baseline model and previously proposed DenseNet model, given the same weights count limit. Because our proposed model can strengthen important connection and discard ineffective ones, our model can also be utilized to explain how each feature affect prediction results.  

\vfill\pagebreak
\bibliographystyle{IEEEbib}
\begin{spacing}{0.9}           % 设置行距
\bibliography{strings,refs}
\end{spacing}

\end{document}